\documentclass[fleqn,usenatbib]{mn2e}

\usepackage[T1]{fontenc}
\usepackage{ae,aecompl}
\usepackage{comment}
\usepackage{graphicx}
\usepackage{multirow}

\usepackage{amsmath,amssymb,graphicx,color,xcolor,longtable,url,enumitem}

\title[Lens potentials with uncertainty-aware DNNs]{Quantifying the structure of strong gravitational lens potentials with uncertainty-aware deep neural networks}

\author[G. Vernardos et al.]{
Georgios Vernardos,$^{1}$\thanks{E-mail: gvernard@ia.forth.gr}
Grigorios Tsagkatakis,$^{2}$
and Yannis Pantazis$^{3}$
\\
$^{1}$Institute of Astrophysics, Foundation for Research and Technology - Hellas (FORTH), GR-70013, Heraklion, Greece \\
$^{2}$Institute of Computer Science, FORTH, GR-70013, Heraklion, Greece \\
$^{3}$Institute of Applied and Computational Mathematics, FORTH, GR-70013, Heraklion, Greece \\
}

\date{Accepted XXX. Received YYY; in original form ZZZ}

\pubyear{2020}

\begin{document}
\label{firstpage}
\pagerange{\pageref{firstpage}--\pageref{lastpage}}
\maketitle

\begin{abstract}
Gravitational lensing is a powerful tool for constraining substructure in the mass distribution of galaxies, be it from the presence of dark matter sub-halos or due to physical mechanisms affecting the baryons throughout galaxy evolution.
Such substructure is hard to model and is either ignored by traditional, smooth modelling, approaches, or treated as well-localized massive perturbers.
In this work, we propose a deep learning approach to quantify the statistical properties of such perturbations directly from images, where only the extended lensed source features within a mask are considered, without the need of any lens modelling.
Our training data consist of mock lensed images assuming perturbing Gaussian Random Fields permeating the smooth overall lens potential, and, for the first time, using images of real galaxies as the lensed source.
We employ a novel deep neural network that can handle arbitrary uncertainty intervals associated with the training dataset labels as input, provides probability distributions as output, and adopts a composite loss function.
The method succeeds not only in accurately estimating the actual parameter values, but also reduces the predicted confidence intervals by 10 per cent in an unsupervised manner, i.e., without having access to the actual ground truth values.
Our results are invariant to the inherent degeneracy between mass perturbations in the lens and complex brightness profiles for the source.
Hence, we can quantitatively and robustly quantify the smoothness of the mass density of thousands of lenses, including confidence intervals, and provide a consistent ranking for follow-up science.
\end{abstract}


\section{Introduction}
Dark matter is a yet undetected component of the Universe, which does not emit light but participates via gravity in the collapse of matter to form galaxies and stars throughout its history \citep{White1978}.
The predominant theory of Cold Dark Matter (CDM), together with dark energy, are successful in explaining the energy density, expansion history, and the large scale ($>$100 Mpc) structure of the Universe \citep{Komatsu2011,PlanckCollaboration2018vi} - the distribution of galaxies into clusters and superclusters.
However, it is on the small, galactic scales ($<$1 Mpc) that different dark matter models begin to diverge \citep{Bullock2017}.
In addition, the dark matter imprints are masked by its interplay with baryons and the several not well-understood non-linear physical mechanisms that become significant as the density rises \citep{Buckley2018}.

Strong gravitational lensing is a powerful technique for detecting dark sub-halos and disentangling the baryonic from the dark mass components in galaxies to probe galaxy formation and evolution in the early Universe \citep{Treu2010}.
The phenomenon involves a distant galaxy-source and a closer galaxy-lens along the line of sight, which deflects the incoming light rays and forms multiple source images, arcs, and rings.
By modelling these features one can assess the overall shape of the lens potential and its smoothness, which are directly linked to underlying dark matter properties (e.g. through the abundance of subhalos) and galaxy evolution/morphology [e.g. through composite lens potentials \citep{Millon2020b}, or higher order moments in the lens mass distribution \citep{Hsueh2017}].
Currently, the total mass distribution in massive elliptical galaxies has been found to be very close to isothermal \citep{Koopmans2006,Koopmans2009,Gavazzi2007,Auger2010,Barnabe2011,Sonnenfeld2013,Suyu2014,Oldham2018}, and the presence of compact, massive, and dark substructures of the order of $10^8$ M$_{\odot}$ has been detected \citep{Vegetti2010,Vegetti2012,Fadely2012,MacLeod2013,Nierenberg2014,Li2016,Hezaveh2016b,Birrer2017}.

Traditional lens modelling techniques attempt to solve the non-linear inverse problem of reconstructing the lensed source brightness and lens potential by optimizing the solution of the lens equation.
This requires pristine data to work with - often complemented with ancillary data, like wide-field observations, spectroscopy, etc - can rely on specific prior assumptions, and can be computationally very expensive, which is particularly true for modelling perturbed lens potentials \citep{Vegetti2010,Vegetti2012}.
Moreover, considerable effort is needed to pre-process the data and to some extent restrict the large parameter space to explore and its inherent degeneracies.
Hence, so far only a few tens of the known lenses have been analyzed (see \citealt{Bolton2008} and \citealt{Wong2020} for the SLACS and H0LiCOW lens samples respectively, which have the most complete observations so far).
This is about to change as the upcoming Euclid space telescope \citep{Laureijs2011} is predicted to discover tens of thousands of galaxy-galaxy lenses in the next decade \citep{Collett2015}, for which it will provide high resolution observations.
This avalanche of data poses a challenge: we need fast and automated algorithms to assess the smoothness of lens potentials, possibly by-passing the intermediate step of smooth modelling, and provide a precise and robust ranking to be considered in allocating resources for in-depth analysis and follow-up observations.

The advent of the Big Data paradigm \citep{FanHanLiu2014} and deep neural network (DNN) algorithms has successfully addressed a number of non-linear image processing problems \citep[e.g.][]{He2016} and is being increasingly used in astrophysics data analysis \citep{Fluke2020}.
Such machine learning techniques employ a representative pre-defined set of observations to train models that can extract information automatically from raw data, successfully mapping the often intractable non-linear parameter space and its degeneracies.
In lensing, DNNs are becoming the mainstream approach to finding lenses \citep[e.g.][]{Metcalf2019}, while they have also been used in lens mass model parameter estimation \citep{Hezaveh2017,Pearson2019} and source reconstruction \citep{Morningstar2019,Chianese2020,Madireddy2020}.

Recently, there has been a boom in using DNNs to study dark matter substructure in lenses.
\citet{DiazRivero2020} approach this as a classification problem, distinguishing between one or more dark CDM subhalos present in a lens.
\citet{Varma2020} use a multi-class classifier to determine the lower mass cut-off of a CDM subhalo population, while \citet{Alexander2020} attempt to distinguish CDM from superfluid dark matter.
\citet{Brehmer2019} adopt a simulation-based inference technique, which allows them to constrain the hyper-parameters of a population of substructures, similarly to a regression problem.
Although the latter approach is the only one that estimates parameters of the perturbing dark matter with confidence intervals, it is a proof-of-principle application and has a quite restricted training set.
A major degeneracy lies between source brightness and lens potential perturbations \citep[see eq. 6 of][]{Koopmans2005} that all of these studies address insufficiently by using a single, or a few, analytic and relatively smooth S\'{e}rsic profiles as source brightness components.
Real lensed sources can include bright star-forming clumps, spiral arms, streams, etc, which when lensed can mimick the effect of a perturbing substructure field.
Moreover, they are based on specific dark matter models and do not examine any baryonic effects that can produce similar lens potential perturbations.

In this work, we present a deep neural network approach to robustly quantify potential perturbations in strong gravitational lenses.
We do not make any specific assumption about underlying dark matter, instead we treat the perturbations as Gaussian Random Fields and measure their statistical properties through their power spectrum.
This allows for substructure due to baryons to be measured as well.
A major novelty of this work is that we consider uncertainties in the training data labels and propose a novel loss function that estimates the probability distribution of each input parameter, thus obtaining a confidence interval for each output value.
The proposed DNN framework marks a clear departure from traditional supervised learning schemes, where training examples are associated with well-defined classes or values and the model is evaluated in terms of its ability to make accurate prediction in new examples coming from the validation set.
In this work, neither the training nor the validation examples are associated with specific values, but rather with distributions containing the corresponding `ground truth' values.
As a result, the DNN model does not have access to these specific values during training, which renders the use of traditional loss function metrics, like cross entropy, not optimal in our case. 

In Section \ref{sec:method} we present the training data and labels, as well as our DNN algorithm.
Section \ref{sec:results} presents our results of training the algorithm on the data.
Our discussion, conclusions, and future prospects are presented in Section \ref{sec:conclusions}.

\section{Methods}
\label{sec:method}
The typical goal of a Machine Learning (ML) model is to learn to assign a label or a value, for classification and regression tasks respectively, to a target sample dataset, as accurately as possible.
The assumption is that each sample is \textit{identifiable}, meaning that it contains sufficient information to make a correct prediction, while errors come from the training inefficiencies or the model's restricted capacity.
In applications where unidentifiability is present, the standard ML interpretation is not suitable because the same sample can be assigned to multiple output values.
To this end, a number of different approaches have been recently presented targeting shortcomings like the existence of noisy labels \citep{ning2019optimization}, partially available labels \citep{durand2019learning} or `soft-labels', where the average of different individual human annotations is utilized as the ground truth \citep{zhang2018mixup}.
These methods attempt to improve the ML model predictions by either correcting the labeling error or by generating artificial training examples, in all cases under a typical supervised learning framework.

In this work, we depart from the usual supervised learning setting and aim to learn the mapping between the lens image and the \textit{distribution} of the target parameter instead of just the values.
However, these distributions are a priori unknown.
We propose to assign artificial, and to some extent arbitrary, distributions to each sample, which correspond to a range of uncertainty for the target parameters, and create a flexible algorithm to learn the true underlying ones.
Hence, we train a DNN whose output is a probability vector on a discretized range for each target parameter.

The lack of the true distribution and its substitution with an artificial and essentially ambiguous one, combined with the freedom of the algorithm to learn essentially any shape, may result in the undesired outcome of increasing the uncertainty.
We alleviate this hindrance by introducing two competing terms in the loss function, one for matching the learned distribution to the target distribution and one that constraints its extent, therefore reducing the uncertainty region.
We anticipate that the DNN model will learn wider distributions at those parameter regimes with more unidentifiability and vice versa.

\begin{figure}
	\begin{center}
	\includegraphics[width=0.5\textwidth,height=0.82\textheight,keepaspectratio]{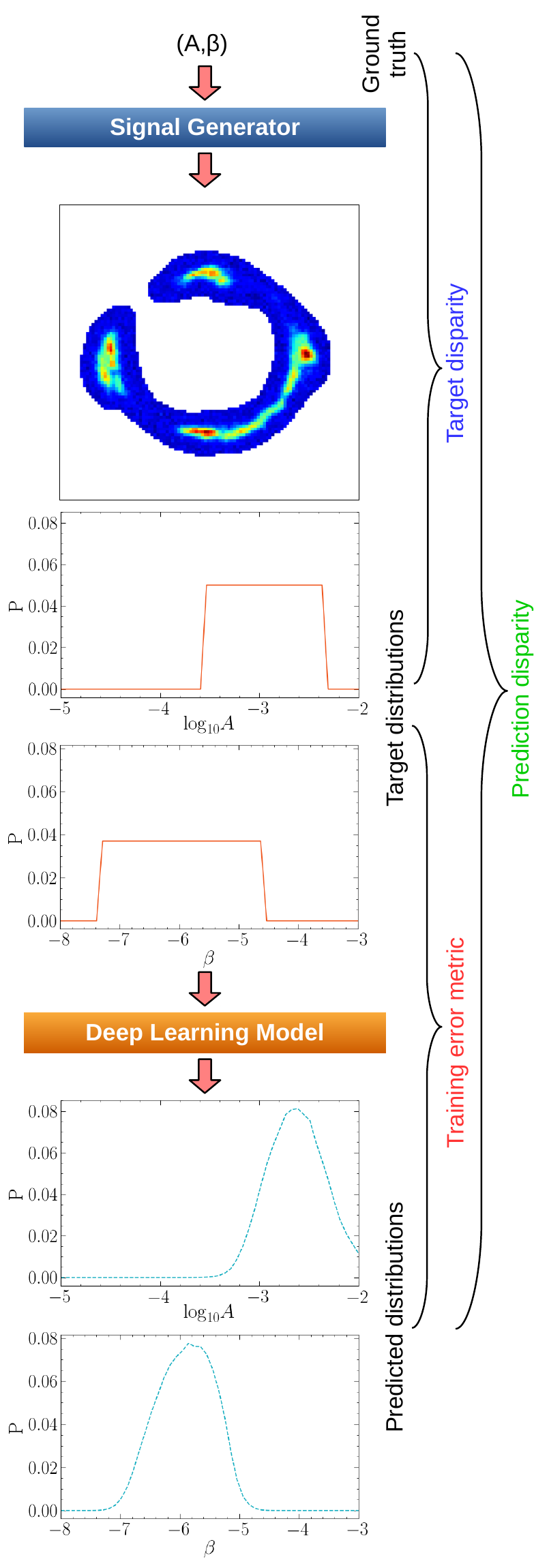}
	\end{center}
	\caption{Visualization of the processing pipeline. Given specific values of the parameters $A$ and $\beta$, the signal generator procedure produces the lens images and associates each image with a distribution of the target parameter values. The DNN model takes the generated images as input and the associated distributions as the target (training labels). The target distribution, however, is associated with value uncertainly (target disparity, defined as the difference between the mean of the target distribution and the ground truth) and the DNN model will accordingly introduce inaccuracies during the prediction (training error metric). In turn, the predicted distribution will also have a different mean and extent (prediction disparity).}
	\label{fig:blockdiagram2}
\end{figure}

A high level overview of the setting considered in this work is presented in Fig. \ref{fig:blockdiagram2}.
Formally, given a set of specific values for the parameters ($A$ and $\beta$, to be introduced in the next section) - the ground truth values - the signal generator produces the corresponding lensed image and the associated parameter distribution, which we call target distribution.
Our DNN model, then, utilizes the generated images as input and provides the predicted distribution with the goal to approximate the target distribution as good as possible.
The error between the target and the predicted distribution is the training error metric, which is employed though the loss function for training the network.
However, the target distribution itself contains uncertainty with respect to the ground truth.
This leads to the existence of the target disparity, which is the difference between the ground truth and parameter estimates derived from the target distribution; here we consider the mean, but other statistical estimators can be used as well.
While the DNN is trained to minimize the training error metric, the ideal behavior would be to minimize the prediction disparity, i.e., the difference between the ground truth values and the mean of the predicted distribution.
In analogy, given a set of images labeled as either an animal or a piece of furniture, the objective would be to recognize whether a given image contains a cat, a dog, a chair, or a desk, namely, without ever having access to images annotated with these specific labels.
In essence, the DNN model never observes the ground truth values during training, which justifies the need to clarify these definitions.

In the first part of this section we present our training sample, with special attention given to creating the training labels, which are defined within an uncertainty interval.
We then describe the Convolutional Neural Network (CNN) architecture that we used, along with the loss function that we employ and our approach in taking into account uncertainty.

\subsection{Training data}
\label{sec:training}
Our training data consists of images of simulated lenses with perturbed lens potentials, viz. an underlying dominant analytic (smooth) model and super-imposed perturbations up to the $\sim15$ per cent level.
Here we describe the different components needed to create such images: the lens potential components, the source brightness profile, and instrumental effects.

\subsubsection{Lens potential}
\label{sec:lens_potential}
For the smooth lens potential, $\psi_{\rm smooth}$, we use the Singular Isothermal Ellipsoid parametric model \citep[SIE,][]{Kassiola1993,Kormann1994} with external shear.
We follow the definition of the SIE convergence given in \citet{Schneider2006}:
\begin{equation}
    \label{eq:kappa}
    \kappa(\omega) = \frac{b}{2\omega},
\end{equation}
where $b$ determines the strength of the potential (the Einstein radius), $\omega=\sqrt{q^2x'^2+y'^2}$, $q$ is the minor to major axis ratio, $x'$ and $y'$ are given in the reference frame of the lens mass center located at $x_0,y_0$ and rotated by the position angle of the ellipsoid on the plane of the sky.
The external shear is defined by its magnitude and direction.
The goal is to create a simulated lens population covering a wide range of physically permitted lenses, without taking into account how representative it is of the real lens population.
Hence, we do not use any observationally motivated priors and the total of 7 free parameters of the smooth model are sampled uniformly within the ranges shown in Table \ref{tab:sampled_params}.

\begin{figure}
	\includegraphics[width=0.5\textwidth]{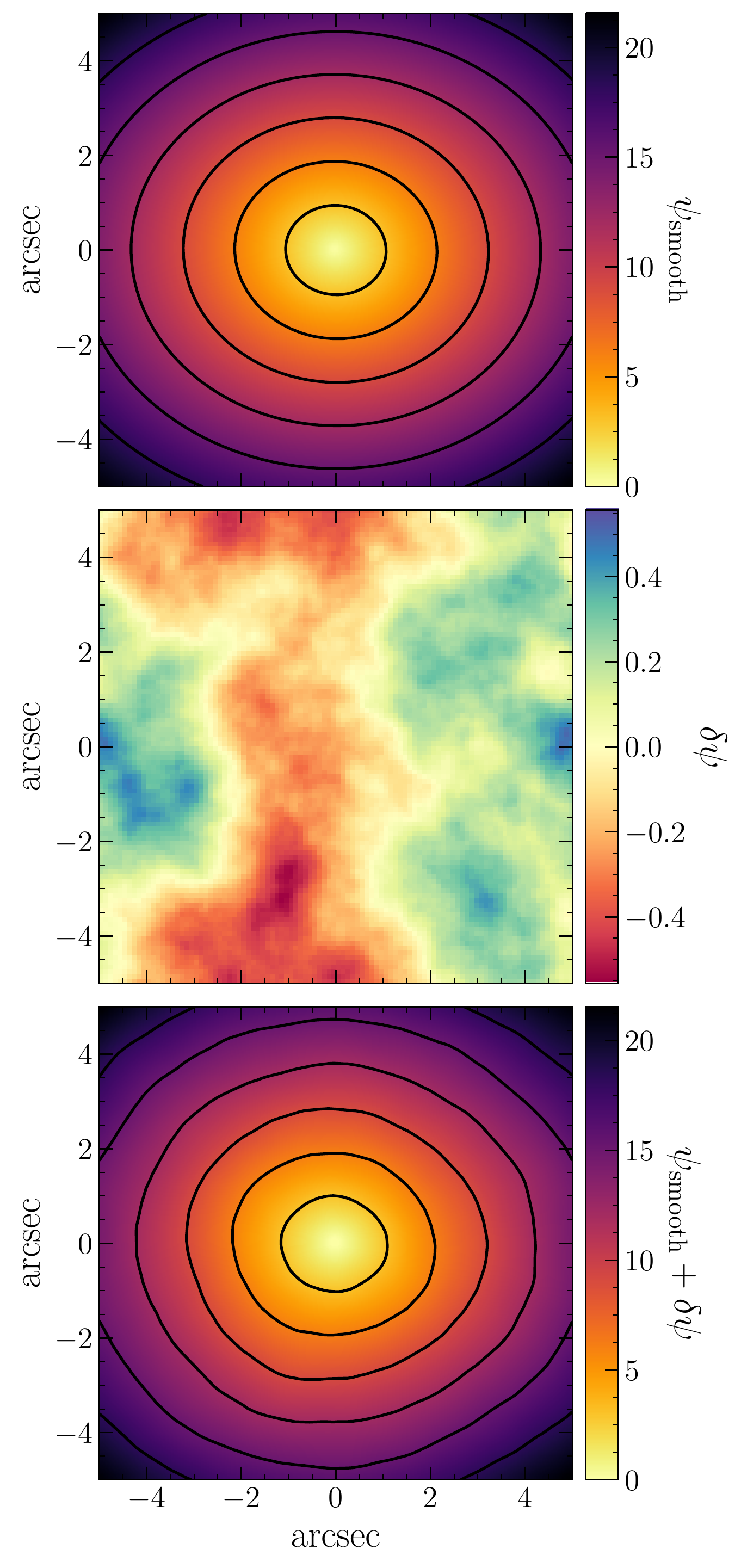}
	\caption{Example of a perturbed lens potential (bottom), composed of a smooth parametric SIE plus external shear model (top) and a realization of GRF perturbations (middle). The smooth model and GRF parameters used in this example are shown in Table \ref{tab:sampled_params}.}
	\label{fig:psi}
\end{figure}

We perturb the smooth potential $\psi_{\rm smooth}$ with Gaussian Random Fields (GRF) of perturbations $\delta\psi$.
GRF perturbations are uniquely defined by a power spectrum, which we assume to be a power law, following \citet{Chatterjee2018,Bayer2018,Chatterjee2019,Vernardos2020a}:
\begin{equation}
\label{eq:power-law}
P(k) = Ak^{\beta},
\end{equation}
where A is the amplitude, associated with the variance of the zero-mean $\delta\psi$ field, $\beta$ is the slope, and $k$ is the wavenumber of the Fourier harmonics.
Regardless of our particular choice of GRF perturbations, the validity of the algorithm presented here is not affected - in fact, any form of potential perturbations could be used.
The range of the $A,\beta$ parameter space that we examine is shown in Table \ref{tab:sampled_params}, from which we draw 2500 random pairs of values.
Each pair of $A$ and $\beta$ values is then used to create a single realization of a corresponding GRF field of lens potential perturbations.
Finally, we combine each perturbation field with a smooth parametric model and create 2500 perturbed lens potentials - an example is shown in Fig. \ref{fig:psi}.

\begin{table*}
	\centering
	\caption{Parameter ranges of the lens potential model composed of a smooth SIE with external shear and super-imposed GRF perturbations defined by a power-law power spectrum. The last column lists the values of the example shown in Fig. \ref{fig:psi}. Angles are defined east-of-north.}
	\label{tab:sampled_params}
	\begin{tabular}{rrrrrr}
	    name        & description               & unit      & min   & max       & example   \\
		\hline
		\multicolumn{5}{l}{SIE with external shear:}\\
		\hline
		$b$         & Einstein radius	        & arcsec    & 1.5   &   3       & 2.56      \\
		$q$         & axis ratio                & -         & 0.5   &   0.999   & 0.71      \\
		$pa$        & position angle            & degrees   & 0     &   180     & 82.53     \\
		$x_0,y_0$   & lens center coordinates   & arcsec    & -0.1  &   0.1     & -0.079,0.042\\
		$\gamma$    & external shear magnitude  & -         & 0     &   0.05    & 0.037     \\
		$\phi$      & external shear direction  & degrees   & 0     &   180     & 14.13     \\
		\hline
		\multicolumn{5}{l}{GRF power-law:}\\
		\hline
		$\mathrm{log_{10}} A$ & power-law amplitude & -   & -5    &   -2      & -2.25     \\
		$\beta$     & power-law slope           & -         & -8    &   -3      & -3.92     \\
		\end{tabular}
\end{table*}

\subsubsection{Source brightness profile}
We use three distinctly different brightness profiles for the lensed source, representing the broad range of galaxies that can be lensed: an analytic profile and archival high resolution observations of two real galaxies, NGC3982 (a spiral galaxy) and NGC2623 (a merger).
The analytic profile consists of two two-dimensional Gaussian components, the first located at $x,y = (-0.05,0.05)$ arcsec on the source plane, with an axis ratio of $0.6$, position angle of $-70^\circ$ (east-of-north), and standard deviation on the $x$ axis of $\sigma_{\rm x} = 0.1$ arcsec, while the second component is at $x,y = (-0.4,0.25)$ arcsec and has $\sigma_{\rm x} = \sigma_{\rm y}= 0.1$ arcsec (symmetric).
The two components are scaled to have a peak brightness ratio of 0.7.
Our choice of extended Gaussian profiles, as opposed to more realistic and concentrated S\'{e}rsic profiles (a common choice in the literature), leads to radially thicker lensed images and serves in creating a broader, more feature-rich training set.
For NGC3982 and NGC2623, we use high resolution archival observations taken with the ACS instrument onboard the Hubble Space Telescope (HST), scaled to roughly much the $\approx1$ arcsec extent of the analytical profile and having the same peak brightness.
The high resolution image that we use for each source is shown in the first column of Fig. \ref{fig:src_img_res}.

\begin{figure}
	\includegraphics[width=0.5\textwidth]{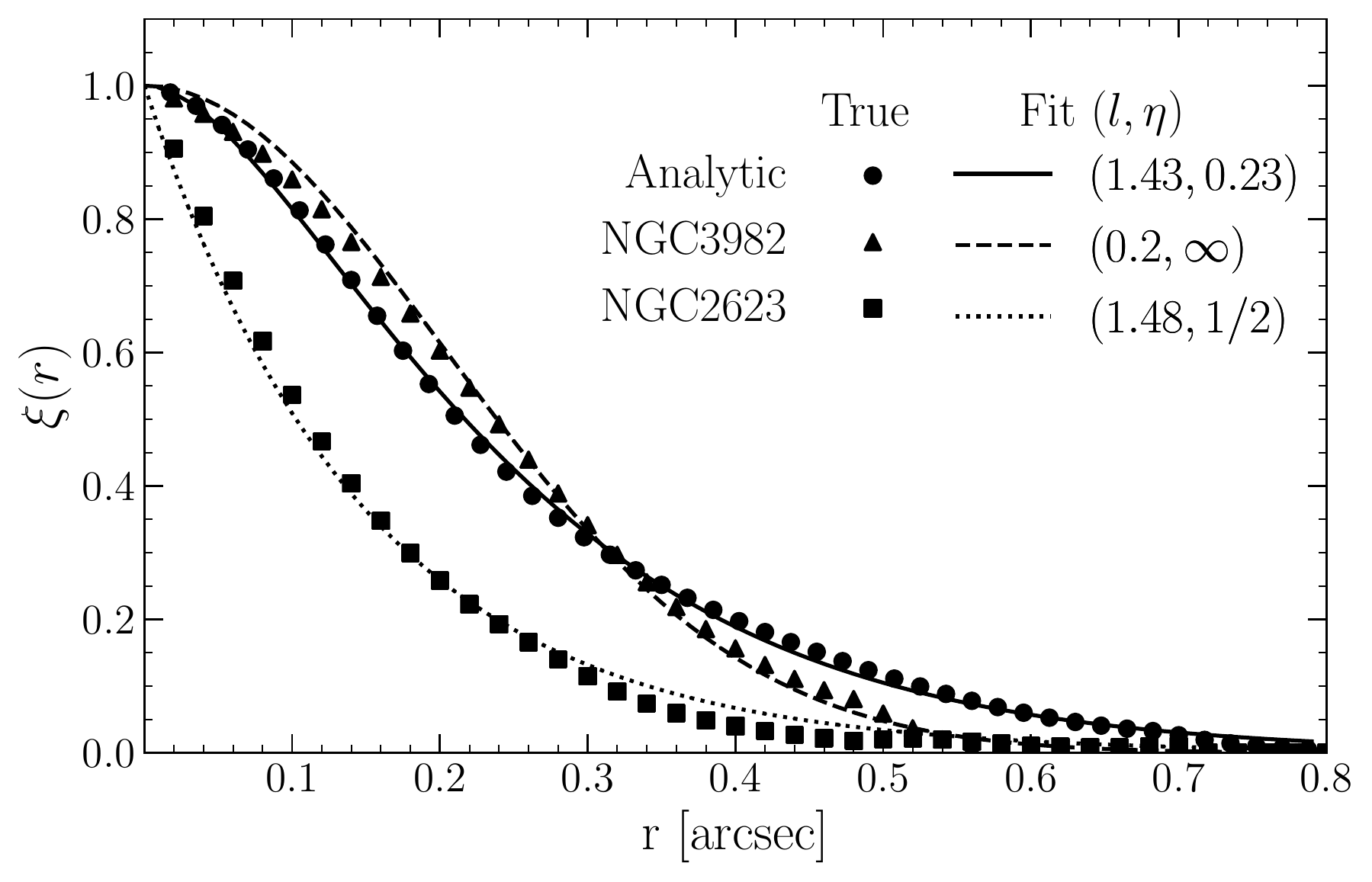}
	\caption{Radially averaged two-point correlation functions for the brightness profiles used as sources in this work, i.e. an analytic profile, NGC3982, and NGC2623, which are shown in the top row of Fig. \ref{fig:src_img_res}. Equation (\ref{eq:matern}) was used to fit the data, with the special cases of $\eta=1/2$ and $\eta \rightarrow \infty$, which correspond to an exponential and a Gaussian, used for NGC2623 and NGC3982 respectively.}
	\label{fig:two_point}
\end{figure}

These sources have distinct covariance properties described accurately by Mat\'{e}rn kernels \citep[][]{Stein1999}:
\begin{equation}
\label{eq:matern}
C(d|l,\eta) = \frac{2^{1-\eta}}{\Gamma(\eta)} \left( \frac{\sqrt{2\eta}d}{l} \right)^{\eta} K_{\rm \eta} \left( \frac{\sqrt{2\eta}d}{l} \right),
\end{equation}
where $d$ is the distance between any two points of the source, $K_{\rm \eta}$ is the modified Bessel function of the second kind, and $l$ is a characteristic coherence scale.
Special cases occur for $\eta=1/2$ and $\eta \rightarrow \infty$, i.e. an exponential and Gaussian kernel, which is the case for NGC2623 and NGC3982 respectively.
The radially averaged source brightness covariance kernels (the two-point correlation function) for these three sources and the values of the best-fit $l$ and $\eta$ from eq. (\ref{eq:matern}) are shown in Fig. \ref{fig:two_point}.

\subsubsection{Simulated lensed images}
We use the \textsc{MOLET}\footnote{\url{https://github.com/gvernard/molet}} software package to generate the mock lenses training dataset \citep[][submitted]{Vernardos2020b}.
For the Point Spread Function (PSF), we generate one for HST using the \textsc{tiny-tim}\footnote{\url{http://www.stsci.edu/hst/observatory/focus/TinyTim}} software \citep{Krist2010}.
Uniform Gaussian random noise with a signal-to-noise ratio of $\approx$40 at peak brightness of the smooth model is added\footnote{\citet{DiazRivero2020} find a 10-20 per cent drop of the accuracy of their classification with respect to adding correlated noise. This, however, is a purely instrumental effect that can be controlled. Hence, without loss of generality, here we adopt uncorrelated noise.}.
The data is simulated on a square 10-arcsec 100-pixel field of view, having a pixel size roughly twice the size of a real HST observation, closer to Euclid's pixel resolution.
Our DNN below is independent of the dimensions of the pixellated images (scale-free), however, the signal-to-noise ratio is encoded in the trained algorithm.
Therefore, although any other image of a lens can be scaled to match the used dimensions (and consequently modify the scale dependent results, viz. the GRF parameters), its noise level has to be the same as the one used here.

\begin{figure*}
	\includegraphics[width=\textwidth]{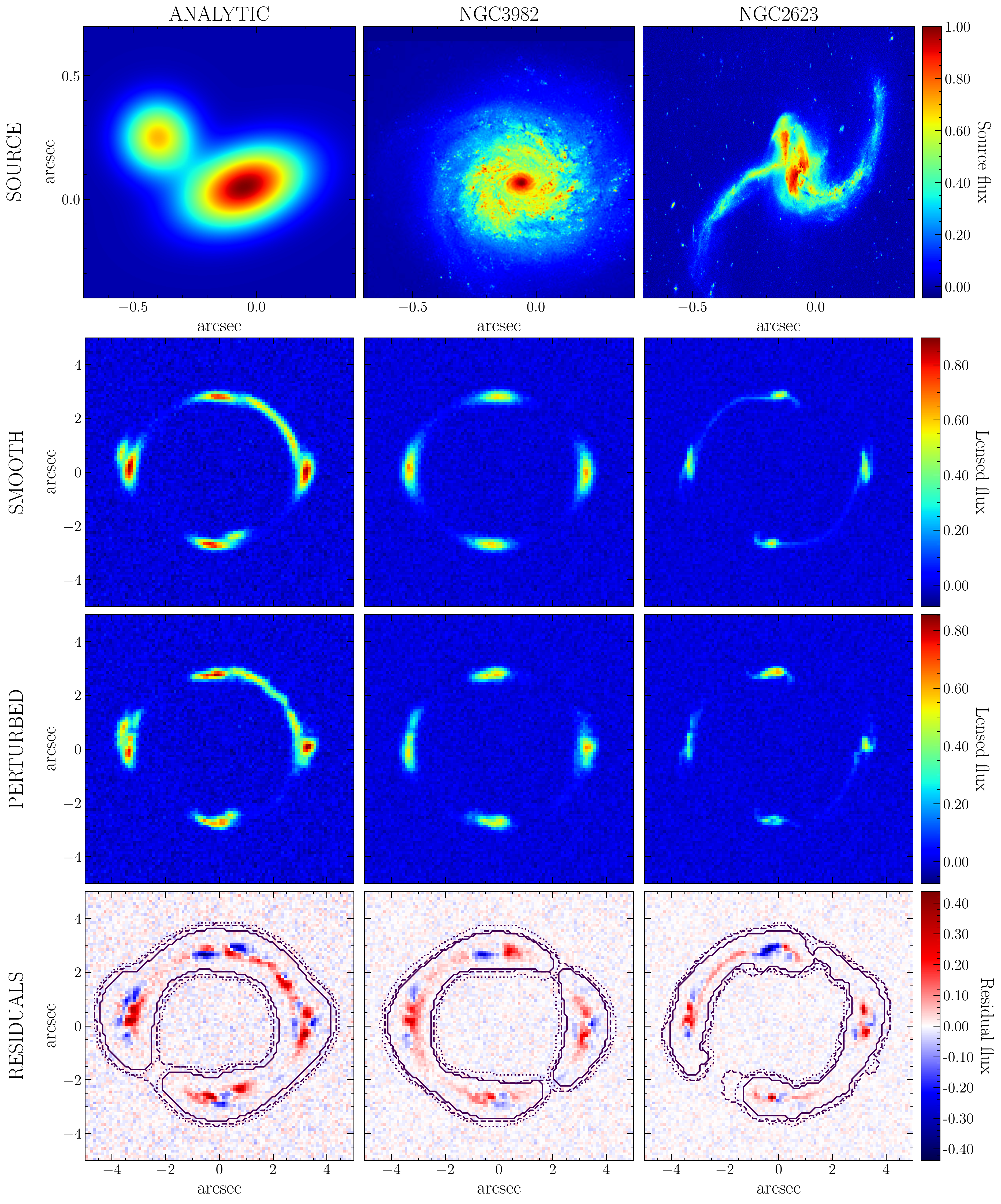}
	\caption{Illustration of the effect of perturbed lens potentials: the three distinct sources that we used (top row), lensed images from smooth (second row) and perturbed (third row) potentials, and the corresponding residuals (bottom row) together with 3 different masks (shown in the residual images for more clarity). The smooth potential and the perturbations are the same for all the examples presented here, shown at the top and middle panels of Fig. \ref{fig:psi}. The data used to train our DNN are the perturbed images (third row), combined with the masks that are indicated by the contours (bottom row).}
	\label{fig:src_img_res}
\end{figure*}

We do not include any light from the lensing galaxy, instead, we mask the field of view away from the lensed source flux.
We use different masks, which allows us to generalize over the mask shape and augment our training dataset.
The masks are generated automatically from the noise-free PSF-convolved lensed images of the smooth lens models.
First, every pixel that has a flux below some threshold, defined as a percentage of the maximum flux, is set to zero.
Then, the image is convolved with a Gaussian filter truncated at the 3$\sigma$ level, where $\sigma$ is a fraction of the radial extent of the lensed images that is determined by finding the maximum inner and minimum outer radius of an annulus encompassing the lensed source flux.
Here, we have created 3 masks per mock lens as a function of two parameters, the threshold and the $\sigma$ of the Gaussian - some examples are shown at the bottom row of Fig. \ref{fig:src_img_res}.

Finally, the perturbed lensed images - shown in the third row of Fig. \ref{fig:src_img_res} - are combined with the 3 different masks and 4 different rotations by 90 degrees on the lens plane.
Hence, for each of the 2500 perturbed lens potentials we produce 12 training samples - a total of 30,000 for each source.

\subsubsection{Defining the training labels}
\begin{figure*}
	\includegraphics[width=\textwidth]{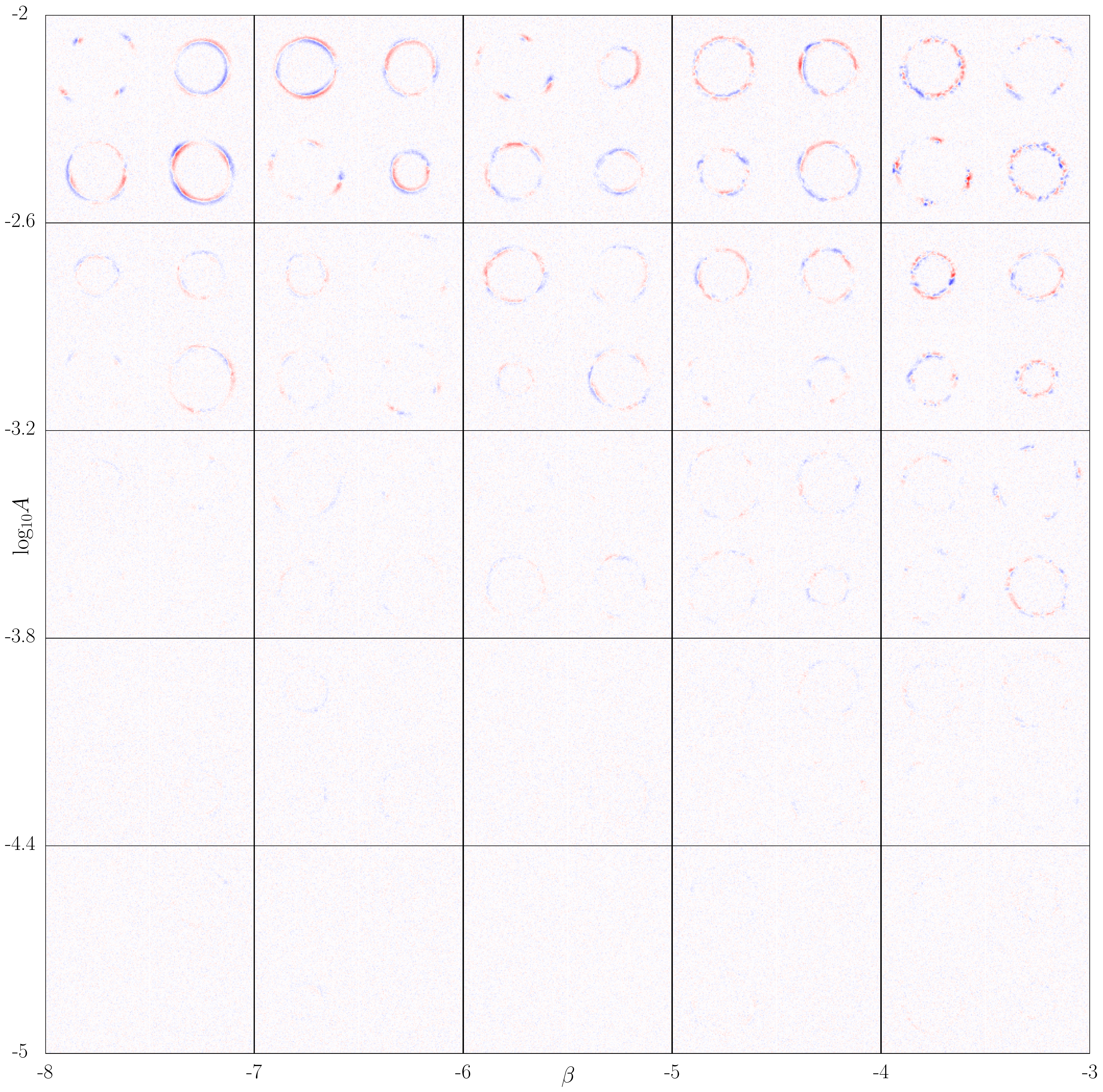}
	\caption{Residual images between simulated lenses with smooth and perturbed lens potentials in the $A,\beta$ parameter space. We arbitrarily divide the parameter space in 25 cells (5$\times$5), drawing 4 random pairs of values from each and showing the corresponding examples. The source is the analytical profile shown in the top left of Fig. \ref{fig:src_img_res}.}
	\label{fig:mosaic}
\end{figure*}

Each image from the mock lens training sample is characterized by two parameters, viz. the GRF power-law amplitude, $A$, and slope, $\beta$.
Assigning uncertainty to the $A,\beta$ of each GRF realization is not straightforward, but to achieve this, one can use the information contained in the power spectrum of residuals between smooth, unperturbed lenses, and their perturbed versions.
Examples of such residuals are shown in Fig. \ref{fig:mosaic} across the $A,\beta$ parameter space.
As anticipated, the main parameter controlling the effect of the perturbations is $A$: for low (high) values of $A$ the power of the signal of the residuals in Fig. \ref{fig:mosaic} is suppressed (enhanced).
The power distribution between the small and large scales, corresponding to large and small wavenumbers $k$ respectively, is determined by $\beta$; high (low) $\beta$ means more (less) power for the small scales.
Therefore, finer structures appear in the residuals as $\beta$ increases in Fig. \ref{fig:mosaic}.

\begin{figure}
	\includegraphics[width=0.5\textwidth]{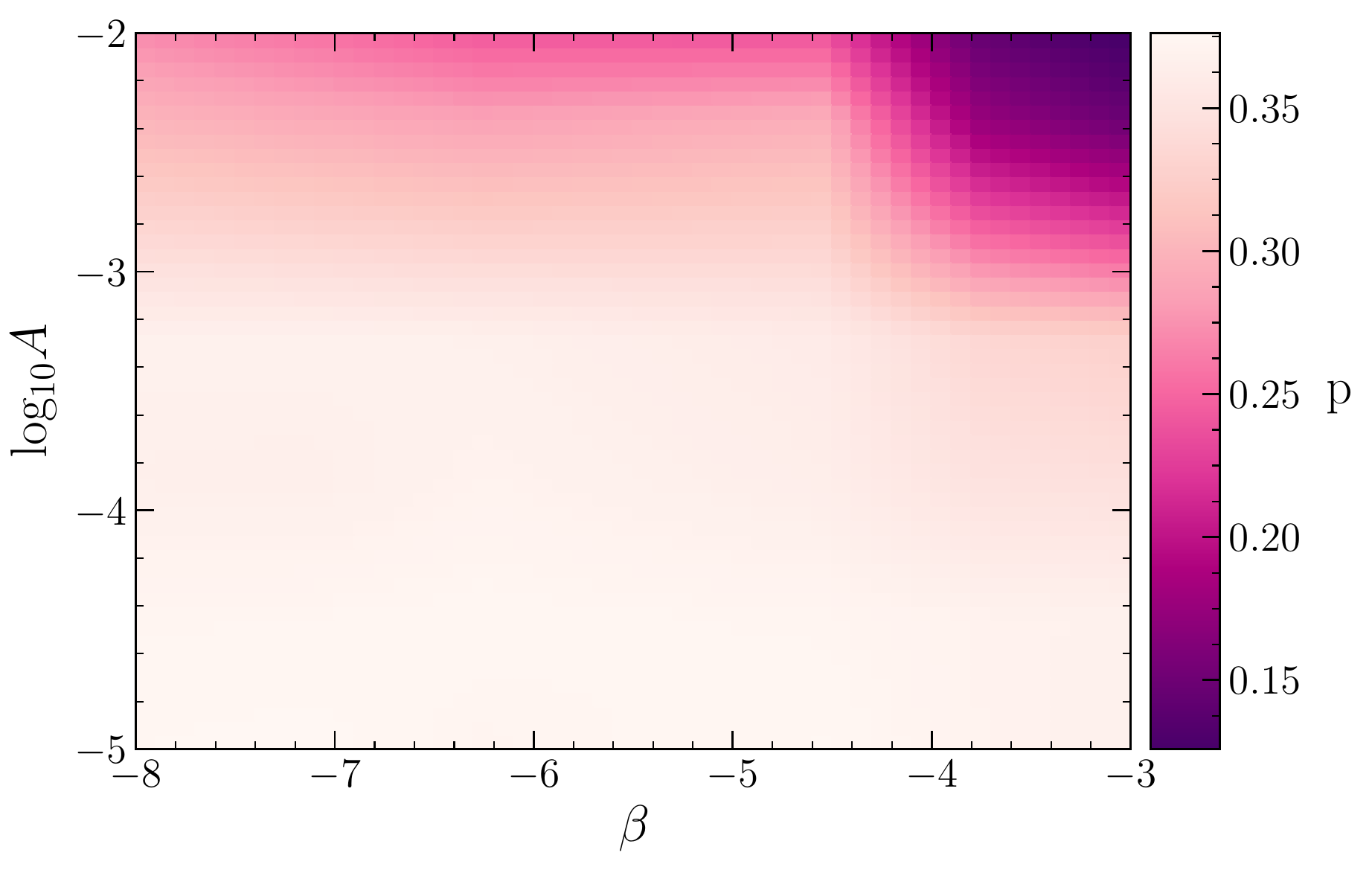}
	\caption{Parameter $p$ of the binomial distribution as a function of $A,\beta$. The values are calculated using equation (\ref{eq:binomial-p}), where $P$ is the average total power in the residual images between perturbed and smooth lens potentials, examples of which are shown in Fig. \ref{fig:mosaic}.}
	\label{fig:prior}
\end{figure}

By calculating the total power of the residuals, $P$, viz. the integral of the power spectrum, one can quantify what is visually perceived in Fig. \ref{fig:mosaic}: for low values of $A$ the residuals are `buried' in the noise and the perturbed lenses look quite similar to the unperturbed ones, while as $A$ and $\beta$ rise the lensing effect of the perturbations becomes increasingly distinct.
We calculate $P$ for 2500 residuals across the $A,\beta$ parameter space (see Fig. \ref{fig:prior}).
The relative values of $P$ (i.e. divided by their maximum) are almost the same, regardless of which of the three sources we choose.

The values of $P$ are scaled and then used to assign a lower and upper bound to each pair of labels $A,\beta$, drawn through a probability distribution.
We want to avoid creating bounds that would lie outside our finite parameter space, therefore we use a binomial distribution (as opposed to a normal distribution).
We create discrete intervals within the given parameter ranges and then sample the number of such intervals used to construct an uncertainty range.
The binomial distribution has two parameters, $N$ and $p$, which we fix through:
\begin{equation}
\label{eq:binomial-N}
N = n N_{\rm p},
\end{equation}
\begin{equation}
\label{eq:binomial-p}
p = \left\{ 1 - \exp \left[ \frac{2P}{\max (P)} \right] \right\}^{-1},
\end{equation}
where we set $n=0.6$ and $N_{\rm p}$ is the number of discrete intervals in the given parameter range; below we set this to 50 but it could be any number.
The appearance of Fig. \ref{fig:prior}, where we show the values of $p$ as a function of $A,\beta$, is matching the one of the residuals shown in Fig. \ref{fig:mosaic}.

\subsection{Deep Neural Network model architecture}
\label{sec:network}

\subsubsection{Problem formulation and network architecture}
Formally, the problem we seek to address is the minimization of the prediction disparity using only the training error metric as the DNN loss function.
To address the specific problem, we employ a CNN in the framework of multi-task regression-via-classification.
Unlike traditional classification settings, where the objective is to predict one class from a set of mutually exclusive classes, here the output class corresponds to a distribution of values within a range. 
Let $\{x_i,z_i \}_{i=1}^{N}$ be the set of $N$ training examples, where each image $x_i$ is associated with $z_i=[log_{10}A_i,\beta_i]$.
In the proposed regression-via-classification setting, instead of predicting the continuous value associated with $z_i$, we form a new target $\hat{z}_i$, where the element $[\hat{A}_i,\hat{\beta}_i]$ produce a set of countable elements given by: 
\begin{equation}
    log_{10}\hat{A} = \Delta \bigg \lfloor \frac{log_{10}A}{\Delta} +\frac{1}{2} \bigg \rfloor  ~~\text{and}~~\\
    \hat{\beta}  = \Delta \bigg \lfloor \frac{\beta}{\Delta} +\frac{1}{2} \bigg \rfloor, \\
\end{equation}
where $\Delta$ is the quantization step size.
The quantization rate, i.e., the range of values grouped into a single value, can be either defined by setting $\Delta$ to a specific value, or by defining the number of bins between the minimum and maximum possible values.
In our case, we consider $50$ bins for both $log_{10}A$ and $\beta$ in the ranges $[-5,-2]$ and $[-8,-3]$ respectively. 

In addition to formulating the problem as an instance of regression-through-classification, the proposed CNN also adheres to the multi-task learning paradigm, where not one but two outputs, one for each parameter, are simultaneously estimated.
The benefit of considering such an approach is that both computationally and performance-wise, only a single set of features are extracted for both outputs, thus implicitly considering the potential feature-level interactions between the two outputs.   

\begin{figure}
	\includegraphics[width=0.5\textwidth]{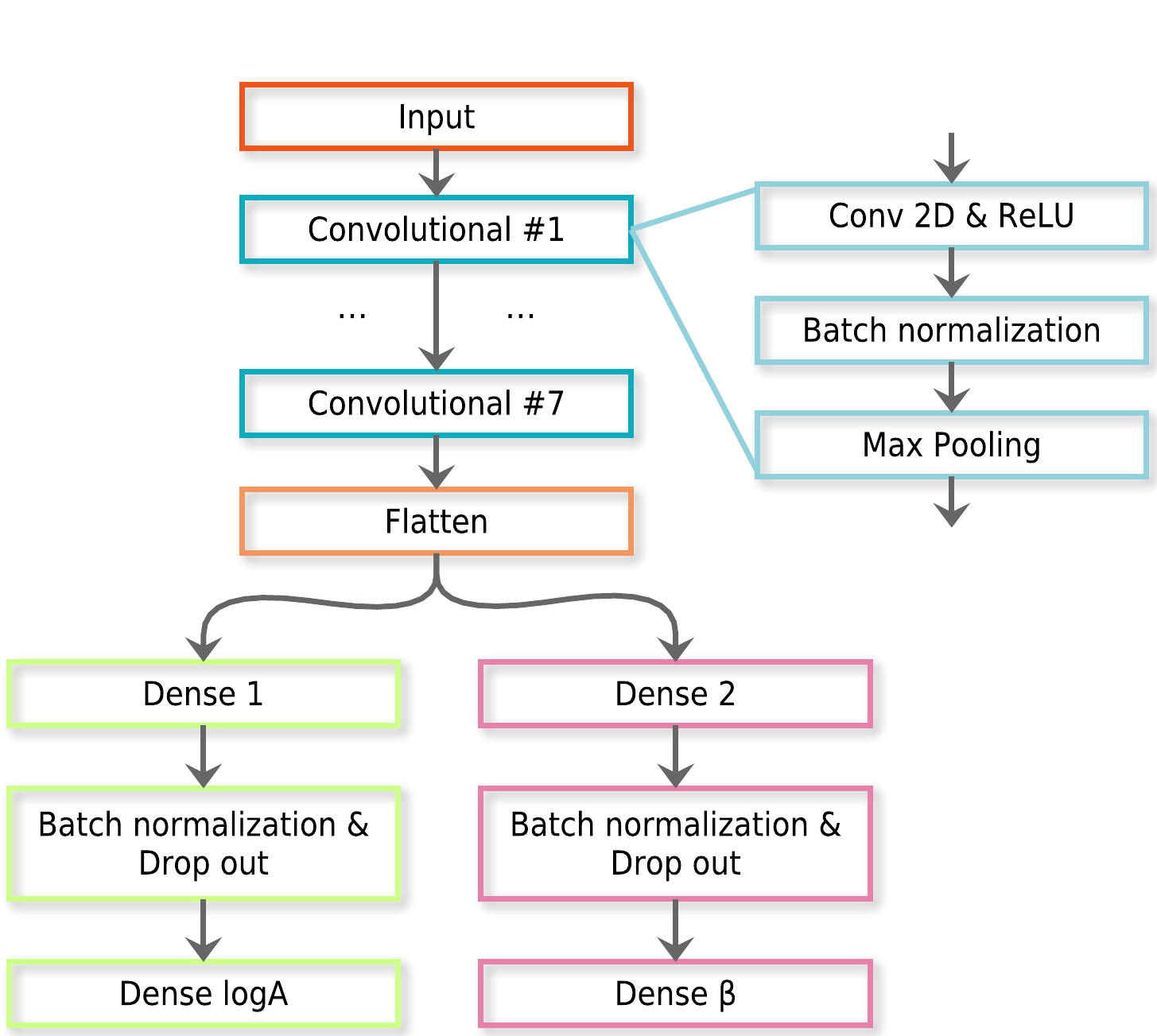}
	\caption{Block diagram visualization of the proposed CNN model architecture. The input image is propagated through a sequence of 7 convolutional blocks, after which it is vectorized (flattened) and split into two streams, one responsible for predicting the value of $\mathrm{log}_{10}A$ and one for predicting the value of $\beta$.}
	\label{fig:blockdiagram}
\end{figure}

Our CNN architecture is outlined in Fig. \ref{fig:blockdiagram}.
We consider an architecture that consists of seven convolutional blocks, whose output is branched to two distinct fully connected blocks, responsible for predicting $A$ and $\beta$ respectively.
Each convolutional block takes as input the output of the previous block, except for the first one that takes the lens image, and applies a two-dimensional convolution, followed by a non-linear ReLU activation, a batch normalization and a max pooling layer.
The output of the last convolutional block is flattened, i.e., converted into a vector, and then passed on to the two fully connected blocks.
Each fully connected block consists of a fully connected layer, a batch normalization layer, a dropout and a final output fully connected layer. 
In our setup, each convolutional layer contains $64$ filters with a spatial range of $3 \times 3$.
Between the flattened layer and each fully connected layer, we additionally introduce a dropout layer with rate $0.5$.
The first fully connected layers (dense $1$ and $2$ in Fig. \ref{fig:blockdiagram}) contain $128$ units and the output layers (Dense $A$ and $\beta$) contain $50$ units, equal to the number of bins.

\subsubsection{Loss function}
In traditional supervised learning, each example is associated with a specific class label and the objective of the ML model is to accurately predict it.
This is typically achieved by using the `one-hot' representation, where the example is represented by a vector of size equal to the total number of classes, whose values are all zeros except the one corresponding to the ground truth class that is one.
We model each example as a random variable characterized by the discrete probability distribution $Q(x)$ and the prediction by $P(x)$.
Therefore, the loss function must be able to capture the similarity or difference between such discrete representations. 

For traditional supervised learning approaches, one can think of the `one-hot' representation as a discrete approximation to a Dirac function.
In this case, minimizing the prediction error is equivalent to minimizing the categorical cross entropy between the target and predicted distributions, given by:
\begin{equation}
CCE(Q\|P)=-\sum_{x \in \mathcal{X}}Q(x) \log ( P(x) ),
\end{equation}
where $\mathcal{X}$ is the set of possible classes.
Categorical cross entropy is a simplified version of the Kullback-Leibler (KL) divergence between distributions, given by:
\begin{equation}
KL(Q\|P)=\sum_{x \in \mathcal{X}}Q(x) \log \Big( \frac{Q(x)}{P(x)} \Big),
\end{equation}
where the fraction is simplified since the objective function seeks to minimize only the predictions given by $P(x)$. 

While for predicting `one-hot' encoded target variables, categorical cross entropy is an excellent choice, it is not able to handle cases where the target variance is characterized by broader families of distributions, such as Binomial, or discrete uniform distributions, as is the case in this work.
To address this challenge, we employ the Jensen-Shannon (JS) divergence as the CNN loss function.
Given the predicted distribution $P$ and the target distribution $Q$, the Jensen-Shannon divergence is defined by:
\begin{equation}
JS(Q,P)=\frac{1}{2}KL(P\|M) + \frac{1}{2}KL(Q\|M),
\end{equation}
where $M=\frac{1}{2}(P+Q)$.
The JS divergence is a generalization of KL that enjoys certain additional benefits, including symmetry and smoothness, while the output is bounded within $[0,1]$.
Unlike categorical cross-entropy, the JS divergence will seek to minimize the difference between the full range of the predicted and target distributions.
This may lead to a situation where the ML model will try to produce results that are characterized by the same amount of uncertainty as that of the training data. 

One way of addressing this challenge is by considering the entropy of the predicted distribution.
Formally, the entropy of the predicted distribution $H(P)$ is given by:
\begin{equation}
H(P)=-\sum_{x \in \mathcal{X}}P(x) \log ( P(x) ).
\end{equation}
Unlike the case of the JS, or even the KL and the CCE functions, the entropy does not consider the target distribution.
Minimizing the prediction entropy resolves to identifying a non-parametric distribution that has the smallest possible extent over the range of classes.
Left by itself, the entropy will tend to produce distributions close to Dirac functions, which in our case will lead to large prediction errors and confidence intervals too small to have any meaning.

To address both challenges associated with the particular problem, namely predict distributions and not single values and reduce the prediction uncertainly with respect to the training sample, we propose a novel loss function that is an entropy-regularized version of the JS divergence and is given by:
\begin{equation}
\mathcal{L}(P,Q)=\lambda JS(P,Q) + (1-\lambda) H(P),  
\end{equation}
where $\lambda$, and correspondingly $(1-\lambda)$, control the impact of the two terms in the compound loss function.
By combining the JS with the entropy, the proposed loss function is able to provide predictions that are both accurate and characterized by significantly smaller uncertainty compared to the target distribution. 
In general, one requires that the loss function is primarily controlled by the $JS(P,Q)$ term, while the entropy term $H(P)$ is introduced as an auxiliary component meant to make the predicted distribution "sharper".  
As such, in our model, we set $\lambda=0.9$ and keep it fixed throughout the training, leaving different combinations and adaptive schemes to be explored in future work.

\section{Results}
\label{sec:results}
\begin{figure}
\includegraphics[width=0.5\textwidth]{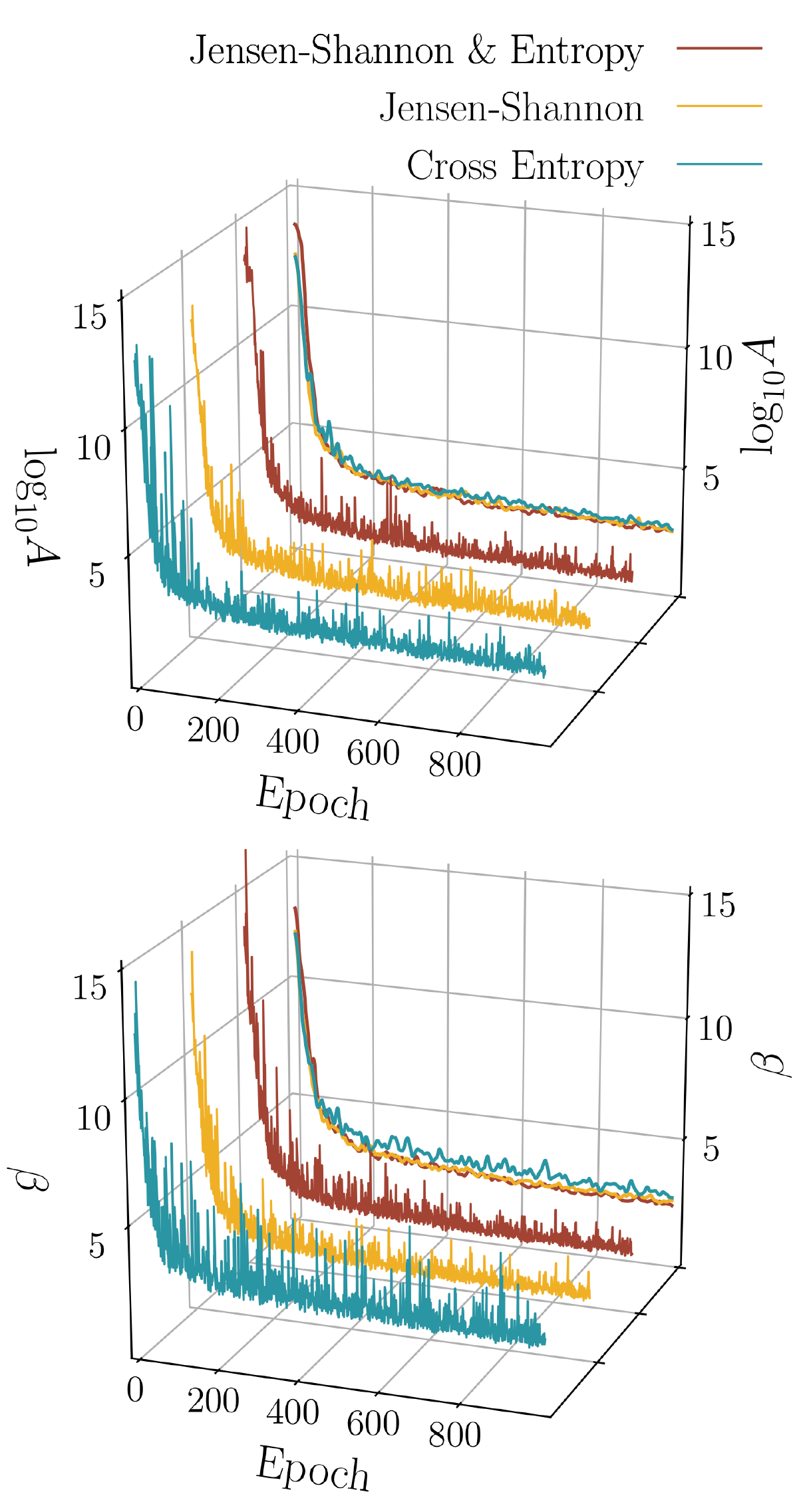}
\caption{Prediction disparity, in units of bins (we used 50 for each parameter in the given range), as a function of training epoch for $\mathrm{log_{10}}A$ (top) and $\beta$ (bottom), for three loss functions: the usual categorical cross entropy, Jensen-Shannon, and the proposed entropy-regularized Jensen-Shannon. The individual results are separated along the y-axis while a smooth average (using a Gaussian filter with $\sigma$=4 epochs) is projected on the $\beta$-epoch plane.}
\label{fig:disparities}
\end{figure}

To train the proposed network, we employ the Adam optimizer \citep{Adam2015}, with the following parameters: learning rate $10^{-3}$, $\beta_1=0.9$, $\beta_2=0.999$ and no learning rate decay, and train it for $1000$ epochs, using $72,000$ examples for training and $18,000$ for validation.
To quantify the performance, we consider the prediction disparity, i.e., the difference between the ground truth parameter values and the mean value of the predicted distribution for the $log_{10}A,\beta$, which is shown in Fig.~\ref{fig:disparities} for the validation examples as a function of epoch.
Overall, we observe that the disparity is reduced as the epochs increase, indicating that the network is able to identify key features and predict the corresponding parameter distributions.

\begin{figure*}
\includegraphics[width=\textwidth]{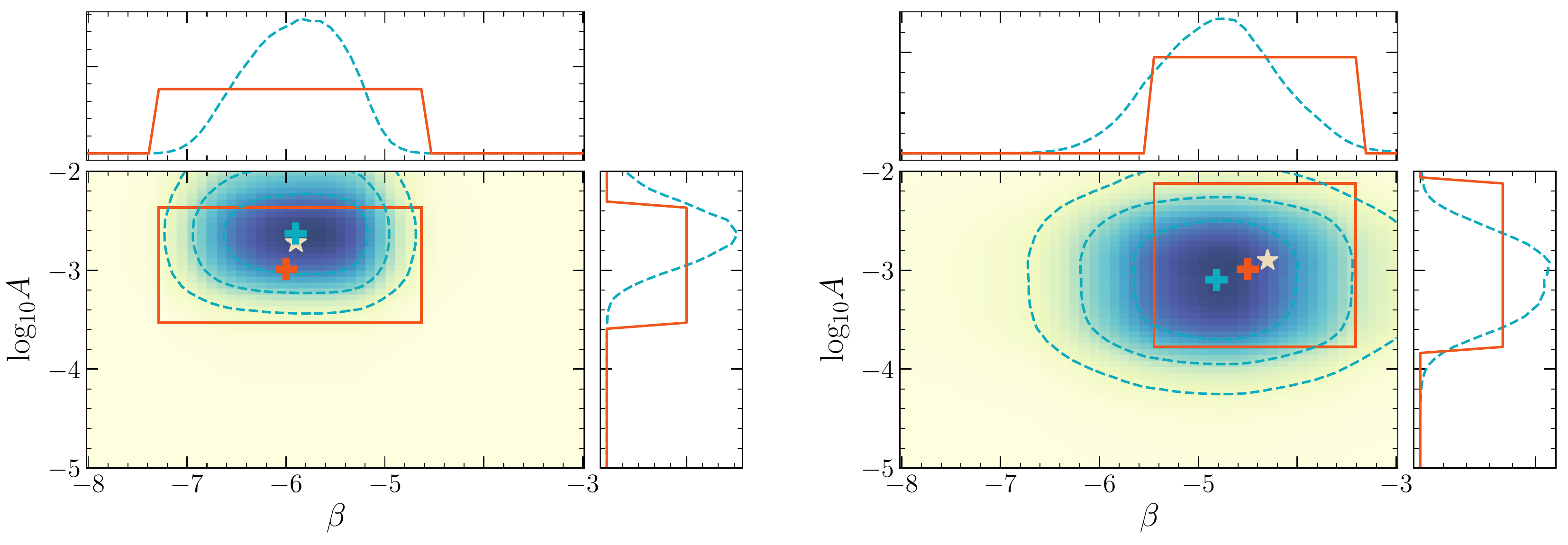}
\caption{Two illustrative examples of the target (solid red) and model prediction (dashed blue) distributions for parameters $A$ and $\beta$. In each panel, we plot the marginal probability distributions of $\mathrm{log}_{10}A$ (side) and $\beta$ (top), and the joint distribution (center). The contours correspond to the 1, 2, and 3 $-\sigma$ intervals respectively, enclosing 68, 95, and 99.7 per cent of the probability volume. The means of the target and predicted distributions are shown (crosses), along with the ground truth value (star). The left panel presents a case where the prediction mean is closer to the ground truth relative to the target mean, while the right panel presents the opposite situation.}
\label{fig:example_results_plot}
\end{figure*}

In Fig. \ref{fig:example_results_plot}, we present two indicative examples of predicted and target distributions, as well as the associated ground truth.
The red rectangle is the uncertainty region encoded into the target distribution, which is in fact the only information that our DNN has access to as training labels.
In both cases, the predicted distribution approximates a Gaussian distribution with most of its mass centered much closer around the mean compared to the uniform distribution of the target.
The left panel shows an example with the predicted distribution being closer to the ground truth than the target, while in the right one the opposite is observed.

Table \ref{tab:disparities} shows the median target and prediction disparities of the two parameters in both training and validation sets.
Unlike traditional supervised learning where the target disparity is zero, i.e., each example is associated with a unique value, in our case, the uncertainty of the target distribution gives rise to disparity for both the training and the validation set.
The results indicate that for the case of the training set, the disparity between model prediction and ground truth is less than the disparity between target and ground truth. Evidently, the model is able to reduce the uncertainty of the training set labels even below the disparity of the target distribution on which it was trained on.
In other words, the trained neural network is capable of balancing between the user-defined uncertainty of the target distribution and the distribution sharpness induced by the entropy penalty in the loss function. However, the case for the validation set is different because the proposed model is required to handle both the uncertainty of the training set labels and the different information contained in the images of the validation sample itself. 

\begin{table}
\centering
\caption{Median prediction and target disparity in units of bins (we used 50 for each parameter in the given range) for the training and validation sets.}
\begin{tabular}{ll|c|c|}
\multicolumn{2}{l}{}                                            & \multicolumn{1}{|c|}{Target}  & \multicolumn{1}{|c|}{Prediction} \\ \hline
\multicolumn{1}{|c|}{\multirow{2}{*}{$A$}}      & Training      & 2.0                           & 1.82                            \\ 
\multicolumn{1}{|c|}{}                          & Validation    & 2.0                           & 2.53                            \\ \hline
\multicolumn{1}{|l|}{\multirow{2}{*}{$\beta$}}  & Training      & 2.0                           & 1.79                            \\  
\multicolumn{1}{|l|}{}                          & Validation    & 2.0                           & 2.47                            \\
\end{tabular}
\label{tab:disparities}
\end{table}

\begin{figure*}
\includegraphics[width=\textwidth]{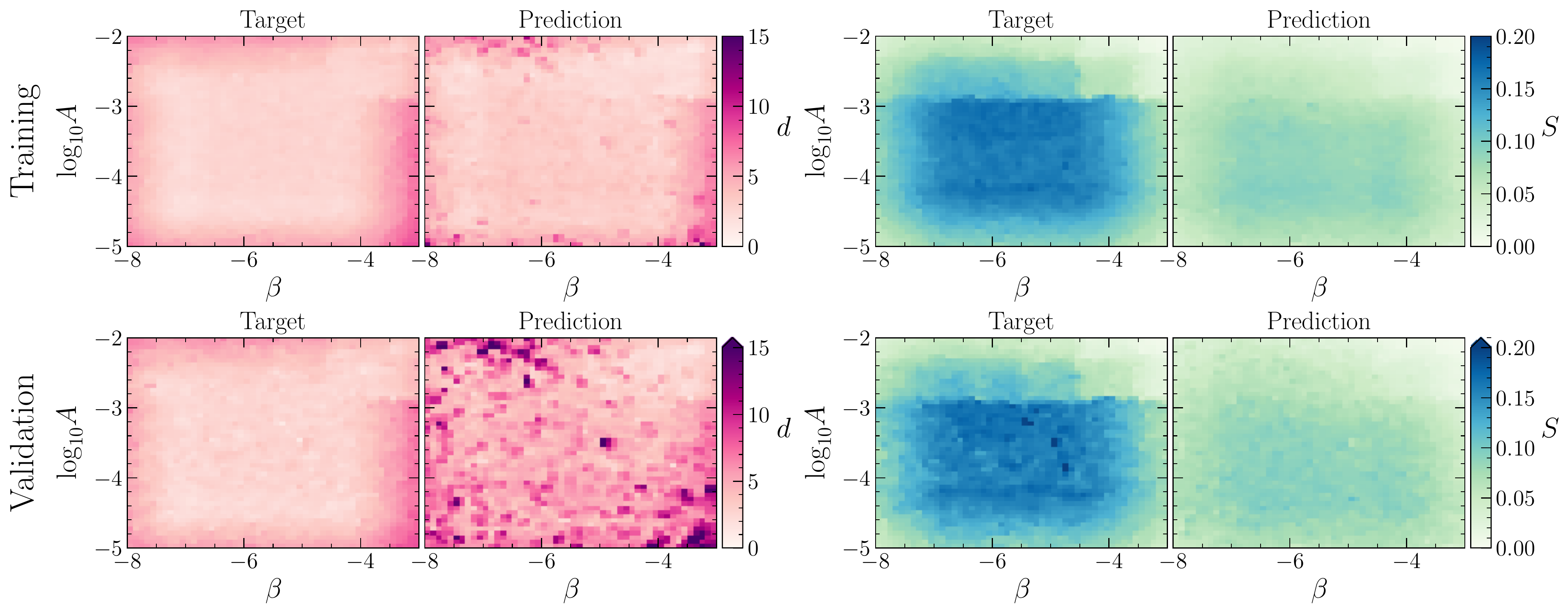}
\caption{Distances, $d$,  (left panels) and areas, $S$,  (right panels) which indicate the accuracy and precision of the prediction and target distributions, as a function of $A,\beta$, for the training (top panels) and validation (bottom panels) sets. The prediction uncertainty corresponds to the $1-\sigma$ area of the distribution (enclosing 68 per cent of the probability), as indicated by the innermost contour in Fig. \ref{fig:example_results_plot}. To get the target uncertainty, we use the confidence intervals on $A,\beta$, assume that the corresponding area indicated by the red rectangle in Fig. \ref{fig:example_results_plot} is in fact a uniform distribution, and calculate the area enclosing 68 per cent of the probability. The distance corresponds to the Euclidean distance of the distribution means and the ground truth, indicated in Fig. \ref{fig:example_results_plot} by the crosses and the star respectively.}
\label{fig:distance_uncertainty}
\end{figure*}

To further analyze the efficiency of the proposed method, we compute two metrics to describe the accuracy and precision of our predictions: i) how close the mean of each distribution is to the ground truth, calculated as the Euclidean distance, $d$, in the $A,\beta$ parameter space (distance between each cross and the star in Fig. \ref{fig:example_results_plot}), and ii) and the area, $S$, covered by each distribution (enclosed by the red rectangle and the $1-\sigma$ contour in Fig. \ref{fig:example_results_plot}).
We compute these metrics for both the target and predicted distributions and present them in Fig. \ref{fig:distance_uncertainty} for both the training and validation sets.
For the training set, we observe that, in general, the distance is very similar for both prediction and target distributions, indicating that our DNN is indeed capable of accurately approximating the mean of the target distribution.
For the validation set, the performance degrades compared to the target distribution, since the model must now handle both the different information contained in the validation set, as well as the uncertainty of the target distribution that is utilized as the training error metric.
Unlike the distance metric, the area, which is directly associated with the prediction uncertainty (precision), is quite better for the predicted distribution compared to the target: in all cases the ground truth is always included in the predicted distribution, which is narrower than the target (although its peak might be offset from the true value, making it less accurate, we underline that this approach is agnostic to the actual $A,\beta$ values used).
This demonstrates that the proposed DNN indeed provides sharper predictions, in the form of confidence intervals, surpassing the target distribution even for the case of the validation set.
In other words, the network is able to mitigate the uncertainty associated with the target distribution under any scenario. 

\section{Discussion and conclusions}
\label{sec:conclusions}
We have created a robust algorithm to quantify lens potential perturbations using just lens images without any lens modelling.
Our approach measures the statistical properties of a perturbing field that pervades the overall lens potential, which is assumed to have a smooth form.
For the first time, we used images of real galaxies as sources that have complex brightness distributions beyond the commonly used analytic profiles, and found that our results are insensitive to the inherent degeneracy between source brightness and lens perturbations.
The resulting method can be easily adapted to specific instrument characteristics, like signal-to-noise-ratio, resolution, seeing, etc, and used to quantify the smoothness of specific existing or future lens samples.
This is important for allocating observational and computational resources to further improve our understanding of dark matter properties, as well as galaxy evolution and interactions.

Our model is not tied to any specific assumptions about the nature of the perturbing field; it can originate from a population of dark matter sub-halos, which can also have different properties for the underlying dark matter particle, or as a result of dynamic baryon-driven galaxy evolution, e.g. mergers.
We specifically examined Gaussian Random Fields, whose power spectrum is exactly defined by a power-law that has two free parameters: the amplitude, $A$, and the slope, $\beta$.
Estimating $A,\beta$ can be thought of as a linear fit to an otherwise unknown power spectrum of different physically motivated perturbing mechanisms.
The presented scheme can, therefore, be used to assess the smoothness of any lens potential to first approximation and without loss of generality.

Here, we used a two-dimensional parametric approach to describe the perturbations, but in reality the physical mechanisms responsible for creating them may be more complicated.
Moreover, we did not explicitly assume any covariance between $A,\beta$, other than an implicit connection through multi-task learning, namely, having the same convolutional layers of our CNN architecture - in Fig. \ref{fig:example_results_plot} we just show the cross product of the one-dimensional distributions that are the output of our neural network.
Increased dimensionality and full parameter covariance can lead to a powerful `free-form' model to describe perturbations, and ultimately dark matter particle properties.
This, together with examining more physically justified models beyond GRFs, is left to be addressed in future work.

In this work, we employ a CNN architecture consisting of $\approx250,000$ parameters, which must be optimized during training from 90,000 examples.
This 2.7 ratio between parameters and training examples could drive the network towards overfitting behavior.
Reducing the size of the network and/or introducing regularization could potentially diminish the presence of such phenomena, as well as produce a smoother convergence during training, as opposed to the spikes observed in Fig. \ref{fig:disparities}.
Another aspect worth investigating is how different ways to generate the target distribution, used as a generalized training label, can affect the performance of the model. 

An important caveat not addressed in this work is the possible contamination of the perturbations' signal from incomplete lens light subtraction.
This is in fact a wider problem in lens modelling \citep{Bolton2006,Marshall2007,Shu2016}; whatever light component exists in the lens plane has to be properly accounted for, either by attributing it to the lens, or to the source, while the perturbations that are of interest here act as a mass component mixing these two under some mass-to-light assumption.
However, a model for the lens light can be treated in the same way as the smooth lens mass model, which is marginalized over in our work.
Conversely, it has been shown that the smooth model parameters, like the Einstein radius, can be successfully recovered by neural networks \citep[e.g.][]{Hezaveh2017,Pearson2019}.
Therefore, although not explicitly examined here, lens light contamination can be easily included in a future extended version of our method.

In this work, we created a homogeneous sample of lensed images assumed to be observed under the same conditions, namely signal-to-noise ratio, noise covariance, seeing (affecting the blurring of the lensed features via the PSF), and instrument resolution.
These properties are indeed expected to be degenerate with lens potential perturbations, especially in the small scales \citep[although noise can become correlated in large scales too, e.g. see][]{DiazRivero2020}, however, their effect is constant across all the different lenses and can be taken into account.
Hence, it is straightforward to re-calibrate our method with a training set created to match the characteristics of a specific instrument, and apply it to assess existing data, e.g. the SLACS \citep{Bolton2008} and BELLS \citep{Shu2016} samples, or make predictions for future instruments, like Euclid.

Finally, we note that in this work, we assume three types of source galaxy profiles, namely, analytic, merger and spiral and train/validate the proposed DNN with examples from all these classes.
This approach offers, to some extent, a source galaxy profile invariance to the trained model.
To truly achieve a profile-independent model, we will explore the adaptation of the proposed learning scheme, where for a given target distribution, the network will be forced to produce the same prediction distribution for all galaxy profiles simultaneously, thus removing any dependency on the specific characteristics of particular profiles. 

The uncertainty-aware DNN-based approach presented in this work, is promising to consistently and robustly rank up to thousands of gravitational lenses in terms of the smoothness of their mass density distribution.
Such a ranking will enable the efficient allocation of observational and computing resources to those most interesting systems, including but not limited to: high resolution imaging, spectroscopy, traditional lens analysis, including galaxy dynamics, population synthesis models for the lens, etc.
Also, performing this analysis can allow for correlations between the derived mass properties and other properties of the lens, such as its type, stellar populations, and environment.
The application of our method looks the most promising for future, large lens samples, like the one expected from Euclid.

\section*{Acknowledgements}
G.V. was funded by the GLADIUS project (contract no. 897124) within the H2020 Framework Program of the European Commission.
G.T. was funded by the CALCHAS project (contract no. 842560) within the H2020 Framework Program of the European Commission.

\section*{Data Availability}
The data that support the findings of this study are available from the corresponding author, G.V., upon reasonable request.

\bibliographystyle{mn2e}
\bibliography{biblio}

\begin{thebibliography}{}

\bibitem[\protect\citeauthoryear{Alexander, Gleyzer, McDonough, Toomey \&
  Usai}{Alexander et~al.}{2020}]{Alexander2020}
Alexander S.,  Gleyzer S.,  McDonough E.,  Toomey M.~W.,    Usai E.,  2020, The
  Astrophysical Journal, 893, 15

\bibitem[\protect\citeauthoryear{Auger, Treu, Bolton, Gavazzi, Koopmans,
  Marshall, Moustakas \& Burles}{Auger et~al.}{2010}]{Auger2010}
Auger M.~W.,  Treu T.,  Bolton A.~S.,  Gavazzi R.,  Koopmans L.~V.,  Marshall
  P.~J.,  Moustakas L.~A.,    Burles S.,  2010, Astrophysical Journal, 724, 511

\bibitem[\protect\citeauthoryear{Barnab{\`{e}}, Czoske, Koopmans, Treu \&
  Bolton}{Barnab{\`{e}} et~al.}{2011}]{Barnabe2011}
Barnab{\`{e}} M.,  Czoske O.,  Koopmans L.~V.,  Treu T.,    Bolton A.~S.,
  2011, Monthly Notices of the Royal Astronomical Society, 415, 2215

\bibitem[\protect\citeauthoryear{Bayer, Chatterjee, Koopmans, Vegetti, McKean,
  Treu \& Fassnacht}{Bayer et~al.}{2018}]{Bayer2018}
Bayer D.,  Chatterjee S.,  Koopmans L. V.~E.,  Vegetti S.,  McKean J.~P.,  Treu
  T.,    Fassnacht C.~D.,  2018, preprint (astro-ph/1803.05952), 23, 1

\bibitem[\protect\citeauthoryear{Birrer \& Amara}{Birrer \&
  Amara}{2017}]{Birrer2017}
Birrer S.,  Amara A.,  2017, Journal of Cosmology and Astroparticle Physics, 5,
  037

\bibitem[\protect\citeauthoryear{Bolton, Burles, Koopmans, Treu, Gavazzi,
  Moustakas, Wayth \& Schlegel}{Bolton et~al.}{2008}]{Bolton2008}
Bolton A.~S.,  Burles S.,  Koopmans L. V.~E.,  Treu T.,  Gavazzi R.,  Moustakas
  L.~A.,  Wayth R.,    Schlegel D.~J.,  2008, The Astrophysical Journal, 682,
  964

\bibitem[\protect\citeauthoryear{Bolton, Burles, Koopmans, Treu \&
  Moustakas}{Bolton et~al.}{2006}]{Bolton2006}
Bolton A.~S.,  Burles S.,  Koopmans L. V.~E.,  Treu T.,    Moustakas L.~a.,
  2006, The Astrophysical Journal, 638, 703

\bibitem[\protect\citeauthoryear{Brehmer, Mishra-Sharma, Hermans, Louppe \&
  Cranmer}{Brehmer et~al.}{2019}]{Brehmer2019}
Brehmer J.,  Mishra-Sharma S.,  Hermans J.,  Louppe G.,    Cranmer K.,  2019,
  The Astrophysical Journal, 886, 49

\bibitem[\protect\citeauthoryear{Buckley \& Peter}{Buckley \&
  Peter}{2018}]{Buckley2018}
Buckley M.~R.,  Peter A.~H.,  2018, Physics Reports, 761, 1

\bibitem[\protect\citeauthoryear{Bullock \& Boylan-Kolchin}{Bullock \&
  Boylan-Kolchin}{2017}]{Bullock2017}
Bullock J.~S.,  Boylan-Kolchin M.,  2017, Annual Review of Astronomy and
  Astrophysics, 55, 343

\bibitem[\protect\citeauthoryear{Chatterjee}{Chatterjee}{2019}]{Chatterjee2019}
Chatterjee S.,  2019, PhD thesis, University of Groningen

\bibitem[\protect\citeauthoryear{Chatterjee \& Koopmans}{Chatterjee \&
  Koopmans}{2018}]{Chatterjee2018}
Chatterjee S.,  Koopmans L.~V.,  2018, Monthly Notices of the Royal
  Astronomical Society, 474, 1762

\bibitem[\protect\citeauthoryear{Chianese, Coogan, Hofma, Otten \&
  Weniger}{Chianese et~al.}{2020}]{Chianese2020}
Chianese M.,  Coogan A.,  Hofma P.,  Otten S.,    Weniger C.,  2020, Monthly
  Notices of the Royal Astronomical Society, 496, 381

\bibitem[\protect\citeauthoryear{Collett}{Collett}{2015}]{Collett2015}
Collett T.,  2015, The Astrophysical Journal, 811, 20

\bibitem[\protect\citeauthoryear{{Diaz Rivero} \& Dvorkin}{{Diaz Rivero} \&
  Dvorkin}{2020}]{DiazRivero2020}
{Diaz Rivero} A.,  Dvorkin C.,  2020, Physical Review D, 101, 1

\bibitem[\protect\citeauthoryear{Durand, Mehrasa \& Mori}{Durand
  et~al.}{2019}]{durand2019learning}
Durand T.,  Mehrasa N.,    Mori G.,  2019, in Proceedings of the IEEE
  Conference on Computer Vision and Pattern Recognition {Learning a deep
  convnet for multi-label classification with partial labels}.
pp 647--657

\bibitem[\protect\citeauthoryear{Fadely \& Keeton}{Fadely \&
  Keeton}{2012}]{Fadely2012}
Fadely R.,  Keeton C.~R.,  2012, Monthly Notices of the Royal Astronomical
  Society, 419, 936

\bibitem[\protect\citeauthoryear{Fan, Han \& Liu}{Fan
  et~al.}{2014}]{FanHanLiu2014}
Fan J.,  Han F.,    Liu H.,  2014, National Science Review, 1, 293

\bibitem[\protect\citeauthoryear{Fluke \& Jacobs}{Fluke \&
  Jacobs}{2020}]{Fluke2020}
Fluke C.~J.,  Jacobs C.,  2020, WIREs Data Mining and Knowledge Discovery, 10,
  e1349

\bibitem[\protect\citeauthoryear{Gavazzi, Treu, Rhodes, Koopmans, Bolton,
  Burles, Massey \& Moustakas}{Gavazzi et~al.}{2007}]{Gavazzi2007}
Gavazzi R.,  Treu T.,  Rhodes J.~D.,  Koopmans V.~E.,  Bolton A.~S.,  Burles
  S.,  Massey R.~J.,    Moustakas L.~A.,  2007, The Astronomical Journal, 667,
  176

\bibitem[\protect\citeauthoryear{He, Zhang, Ren \& Sun}{He
  et~al.}{2016}]{He2016}
He K.,  Zhang X.,  Ren S.,    Sun J.,  2016, in Proceedings of the IEEE
  conference on computer vision and pattern recognition {Deep residual learning
  for image recognition}.
p.~770

\bibitem[\protect\citeauthoryear{Hezaveh, Dalal, Marrone, Mao, Morningstar,
  Wen, Blandford, Carlstrom, Fassnacht, Holder, Kemball, Marshall, Murray,
  Levasseur, Vieira \& Wechsler}{Hezaveh et~al.}{2016}]{Hezaveh2016b}
Hezaveh Y.~D.,  Dalal N.,  Marrone D.~P.,  Mao Y.-Y.,  Morningstar W.,  Wen D.,
   Blandford R.~D.,  Carlstrom J.~E.,  Fassnacht C.~D.,  Holder G.~P.,  Kemball
  A.,  Marshall P.~J.,  Murray N.,  Levasseur L.~P.,  Vieira J.~D.,    Wechsler
  R.~H.,  2016, The Astrophysical Journal, 823, 1

\bibitem[\protect\citeauthoryear{Hezaveh, Levasseur \& Marshall}{Hezaveh
  et~al.}{2017}]{Hezaveh2017}
Hezaveh Y.~D.,  Levasseur L.~P.,    Marshall P.~J.,  2017, Nature, 548, 555

\bibitem[\protect\citeauthoryear{Hsueh, Oldham, Spingola, Vegetti, Fassnacht,
  Auger, Koopmans, McKean \& Lagattuta}{Hsueh et~al.}{2017}]{Hsueh2017}
Hsueh J.~W.,  Oldham L.,  Spingola C.,  Vegetti S.,  Fassnacht C.~D.,  Auger
  M.~W.,  Koopmans L.~V.,  McKean J.~P.,    Lagattuta D.~J.,  2017, Monthly
  Notices of the Royal Astronomical Society, 469, 3713

\bibitem[\protect\citeauthoryear{Kassiola \& Kovner}{Kassiola \&
  Kovner}{1993}]{Kassiola1993}
Kassiola A.,  Kovner I.,  1993, The Astrophysical Journal, 417, 450

\bibitem[\protect\citeauthoryear{Kingma \& Ba}{Kingma \& Ba}{2015}]{Adam2015}
Kingma D.~P.,  Ba J.,  2015, in Bengio Y.,  LeCun Y.,  eds, 3rd International
  Conference on Learning Representations, {\{}ICLR{\}} 2015, San Diego, CA,
  USA, May 7-9, 2015, Conference Track Proceedings {Adam: A Method for
  Stochastic Optimization}

\bibitem[\protect\citeauthoryear{Komatsu, Smith, Dunkley, Bennett, Gold,
  Hinshaw, Jarosik, Larson, Nolta, Page, Spergel, Halpern, Hill, Kogut, Limon,
  Meyer, Odegard, Tucker, Weiland, Wollack \& Wright}{Komatsu
  et~al.}{2011}]{Komatsu2011}
Komatsu E.,  Smith K.~M.,  Dunkley J.,  Bennett C.~L.,  Gold B.,  Hinshaw G.,
  Jarosik N.,  Larson D.,  Nolta M.~R.,  Page L.,  Spergel D.~N.,  Halpern M.,
  Hill R.~S.,  Kogut A.,  Limon M.,  Meyer S.~S.,  Odegard N.,  Tucker G.~S.,
  Weiland J.~L.,  Wollack E.,    Wright E.~L.,  2011, Astrophysical Journal,
  Supplement Series, 192

\bibitem[\protect\citeauthoryear{Koopmans}{Koopmans}{2005}]{Koopmans2005}
Koopmans L.~V.,  2005, Monthly Notices of the Royal Astronomical Society, 363,
  1136

\bibitem[\protect\citeauthoryear{Koopmans, Bolton, Treu, Czoske, Auger,
  Barnab{\`{e}}, Vegetti, Gavazzi, Moustakas \& Burles}{Koopmans
  et~al.}{2009}]{Koopmans2009}
Koopmans L.~V.,  Bolton A.,  Treu T.,  Czoske O.,  Auger M.~W.,  Barnab{\`{e}}
  M.,  Vegetti S.,  Gavazzi R.,  Moustakas L.~A.,    Burles S.,  2009,
  Astrophysical Journal, 703, 51

\bibitem[\protect\citeauthoryear{Koopmans, Treu, Bolton, Burles \&
  Moustakas}{Koopmans et~al.}{2006}]{Koopmans2006}
Koopmans L. V.~E.,  Treu T.,  Bolton A.~S.,  Burles S.,    Moustakas L.~A.,
  2006, The Astrophysical Journal, 649, 599

\bibitem[\protect\citeauthoryear{Kormann, Schneider \& Bartelmann}{Kormann
  et~al.}{1994}]{Kormann1994}
Kormann R.,  Schneider P.,    Bartelmann M.,  1994, Astronomy {\&}
  Astrophysics, 284, 285

\bibitem[\protect\citeauthoryear{Krist, Hook \& Stoehr}{Krist
  et~al.}{2010}]{Krist2010}
Krist J.,  Hook R.,    Stoehr F.,  2010, Astrophysics Source Code Library, p.
  record ascl:1010.057

\bibitem[\protect\citeauthoryear{Laureijs, Amiaux, Arduini, Augu{\`{e}}res,
  Brinchmann, Cole, Cropper, Dabin, Duvet, Ealet, Garilli, Gondoin, Guzzo,
  Hoar, Hoekstra, Holmes, Kitching, Maciaszek \& Mellier}{Laureijs
  et~al.}{2011}]{Laureijs2011}
Laureijs R.,  Amiaux J.,  Arduini S.,  Augu{\`{e}}res J.~L.,  Brinchmann J.,
  Cole R.,  Cropper M.,  Dabin C.,  Duvet L.,  Ealet A.,  Garilli B.,  Gondoin
  P.,  Guzzo L.,  Hoar J.,  Hoekstra H.,  Holmes R.,  Kitching T.,  Maciaszek
  T.,    Mellier Y.,  2011, preprint (astro-ph/1110.3193)

\bibitem[\protect\citeauthoryear{Li, Frenk, Cole, Gao, Bose \& Hellwing}{Li
  et~al.}{2016}]{Li2016}
Li R.,  Frenk C.~S.,  Cole S.,  Gao L.,  Bose S.,    Hellwing W.~A.,  2016,
  Monthly Notices of the Royal Astronomical Society, 460

\bibitem[\protect\citeauthoryear{MacLeod, Jones, Agol \& Kochanek}{MacLeod
  et~al.}{2013}]{MacLeod2013}
MacLeod C.~L.,  Jones R.,  Agol E.,    Kochanek C.~S.,  2013, The Astrophysical
  Journal, 773, 35

\bibitem[\protect\citeauthoryear{Madireddy, Li, Ramachandra, Balaprakash \&
  Habib}{Madireddy et~al.}{2020}]{Madireddy2020}
Madireddy S.,  Li N.,  Ramachandra N.,  Balaprakash P.,    Habib S.,  2020,
  preprint (astro-ph/1911.03867)

\bibitem[\protect\citeauthoryear{Marshall, Treu, Melbourne, Gavazzi, Bundy,
  Ammons, Bolton, Burles, Larkin, {Le Mignant}, Koo, Koopmans, Max, Moustakas,
  Steinbring \& Wright}{Marshall et~al.}{2007}]{Marshall2007}
Marshall P.~J.,  Treu T.,  Melbourne J.,  Gavazzi R.,  Bundy K.,  Ammons S.~M.,
   Bolton A.~S.,  Burles S.,  Larkin J.~E.,  {Le Mignant} D.,  Koo D.~C.,
  Koopmans L. V.~E.,  Max C.~E.,  Moustakas L.~A.,  Steinbring E.,    Wright
  S.~A.,  2007, The Astrophysical Journal, 671, 1196

\bibitem[\protect\citeauthoryear{Metcalf, Meneghetti, Avestruz, Bellagamba,
  Bom, Bertin, Cabanac, Davies, Decenci{\`{e}}re, Flamary, Gavazzi, Geiger,
  Hartley, Huertas-Company \& Jackson}{Metcalf et~al.}{2019}]{Metcalf2019}
Metcalf R.~B.,  Meneghetti M.,  Avestruz C.,  Bellagamba F.,  Bom C.~R.,
  Bertin E.,  Cabanac R.,  Davies A.,  Decenci{\`{e}}re E.,  Flamary R.,
  Gavazzi R.,  Geiger M.,  Hartley P.,  Huertas-Company M.,    Jackson N.,
  2019, Astronomy {\&} Astrophysics, 625, A119

\bibitem[\protect\citeauthoryear{Millon, Galan, Courbin, Treu, Suyu, Ding,
  Birrer, Chen, Shajib, Sluse, Wong, Agnello, Auger, Buckley-Geer, Chan \&
  Collett}{Millon et~al.}{2020}]{Millon2020b}
Millon M.,  Galan A.,  Courbin F.,  Treu T.,  Suyu S.~H.,  Ding X.,  Birrer S.,
   Chen G.~C.,  Shajib A.~J.,  Sluse D.,  Wong K.~C.,  Agnello A.,  Auger
  M.~W.,  Buckley-Geer E.~J.,  Chan J.~H.,    Collett T.,  2020, Astronomy and
  Astrophysics, 639, 1

\bibitem[\protect\citeauthoryear{Morningstar, Levasseur, Hezaveh, Blandford,
  Marshall, Putzky, Rueter, Wechsler \& Welling}{Morningstar
  et~al.}{2019}]{Morningstar2019}
Morningstar W.~R.,  Levasseur L.~P.,  Hezaveh Y.~D.,  Blandford R.,  Marshall
  P.,  Putzky P.,  Rueter T.~D.,  Wechsler R.,    Welling M.,  2019, preprint
  (astro-ph/1901.01359)

\bibitem[\protect\citeauthoryear{Nierenberg, Treu, Wright, Fassnacht \&
  Auger}{Nierenberg et~al.}{2014}]{Nierenberg2014}
Nierenberg A.~M.,  Treu T.,  Wright S.~A.,  Fassnacht C.~D.,    Auger M.~W.,
  2014, Monthly Notices of the Royal Astronomical Society, 442, 2434

\bibitem[\protect\citeauthoryear{Ning \& You}{Ning \&
  You}{2019}]{ning2019optimization}
Ning C.,  You F.,  2019, Computers {\&} Chemical Engineering, 125, 434

\bibitem[\protect\citeauthoryear{Oldham \& Auger}{Oldham \&
  Auger}{2018}]{Oldham2018}
Oldham L.~J.,  Auger M.~W.,  2018, Monthly Notices of the Royal Astronomical
  Society, 476, 133

\bibitem[\protect\citeauthoryear{Pearson, Li \& Dye}{Pearson
  et~al.}{2019}]{Pearson2019}
Pearson J.,  Li N.,    Dye S.,  2019, Monthly Notices of the Royal Astronomical
  Society, 488, 991

\bibitem[\protect\citeauthoryear{{Planck Collaboration}, Aghanim, Akrami,
  Ashdown, Aumont, Baccigalupi, Ballardini, Banday, Barreiro, Bartolo, Basak,
  Battye, Benabed \& Bernard}{{Planck Collaboration}
  et~al.}{2018}]{PlanckCollaboration2018vi}
{Planck Collaboration} P.,  Aghanim N.,  Akrami Y.,  Ashdown M.,  Aumont J.,
  Baccigalupi C.,  Ballardini M.,  Banday A.~J.,  Barreiro R.~B.,  Bartolo N.,
  Basak S.,  Battye R.,  Benabed K.,    Bernard J.~P.,  2018, preprint
  (astro-ph/1807.06209)

\bibitem[\protect\citeauthoryear{Schneider, Kochanek \& Wambsganss}{Schneider
  et~al.}{2006}]{Schneider2006}
Schneider P.,  Kochanek C.~S.,    Wambsganss J.,  2006, in G M.,  P. J.,   P.
  N.,  eds, Saas-Fee Advanced Course vol. 33 {Gravitational Lensing: Strong,
  Weak, Micro}.
Springer, Berlin

\bibitem[\protect\citeauthoryear{Shu, Bolton, Mao, Kochanek,
  P{\'{e}}rez-Fournon, Oguri, Montero-Dorta, Cornachione, Marques-Chaves,
  Zheng, Brownstein \& M{\'{e}}nard}{Shu et~al.}{2016}]{Shu2016}
Shu Y.,  Bolton A.~S.,  Mao S.,  Kochanek C.~S.,  P{\'{e}}rez-Fournon I.,
  Oguri M.,  Montero-Dorta A.~D.,  Cornachione M.~A.,  Marques-Chaves R.,
  Zheng Z.,  Brownstein J.~R.,    M{\'{e}}nard B.,  2016, The Astrophysical
  Journal, 833, 264

\bibitem[\protect\citeauthoryear{Sonnenfeld, Treu, Suyu, Marshall, Auger \&
  Nipoti}{Sonnenfeld et~al.}{2013}]{Sonnenfeld2013}
Sonnenfeld A.,  Treu T.,  Suyu S.~H.,  Marshall P.~J.,  Auger M.~W.,    Nipoti
  C.,  2013, The Astrophysical Journal, 777, 98

\bibitem[\protect\citeauthoryear{Stein}{Stein}{1999}]{Stein1999}
Stein M.,  1999, {Interpolation of Spatial Data: Some Theory for Kriging},
  springer s edn.
Springer, New York

\bibitem[\protect\citeauthoryear{Suyu, Treu, Hilbert, Sonnenfeld, Auger,
  Blandford, Collett, Courbin, Fassnacht, Koopmans, Marshall, Meylan, Spiniello
  \& Tewes}{Suyu et~al.}{2014}]{Suyu2014}
Suyu S.~H.,  Treu T.,  Hilbert S.,  Sonnenfeld A.,  Auger M.~W.,  Blandford
  R.~D.,  Collett T.,  Courbin F.,  Fassnacht C.~D.,  Koopmans L. V.~E.,
  Marshall P.~J.,  Meylan G.,  Spiniello C.,    Tewes M.,  2014, The
  Astrophysical Journal Letters, 788, L35

\bibitem[\protect\citeauthoryear{Treu}{Treu}{2010}]{Treu2010}
Treu T.,  2010, Annu. Rev. of Astron. and Astrophys., 48, 87

\bibitem[\protect\citeauthoryear{Varma, Fairbairn \& Figueroa}{Varma
  et~al.}{2020}]{Varma2020}
Varma S.,  Fairbairn M.,    Figueroa J.,  2020, preprint (astro-ph/2005.05353)

\bibitem[\protect\citeauthoryear{Vegetti, Koopmans, Bolton, Treu \&
  Gavazzi}{Vegetti et~al.}{2010}]{Vegetti2010}
Vegetti S.,  Koopmans L. V.~E.,  Bolton A.,  Treu T.,    Gavazzi R.,  2010,
  Monthly Notices of the Royal Astronomical Society, 408, 1969

\bibitem[\protect\citeauthoryear{Vegetti, Lagattuta, McKean, Auger, Fassnacht
  \& Koopmans}{Vegetti et~al.}{2012}]{Vegetti2012}
Vegetti S.,  Lagattuta D.~J.,  McKean J.~P.,  Auger M.~W.,  Fassnacht C.~D.,
  Koopmans L.~V.,  2012, Nature, 481, 341

\bibitem[\protect\citeauthoryear{Vernardos}{Vernardos}{2020}]{Vernardos2020b}
Vernardos G.,  2020, in preparation

\bibitem[\protect\citeauthoryear{Vernardos \& Koopmans}{Vernardos \&
  Koopmans}{2020}]{Vernardos2020a}
Vernardos G.,  Koopmans L. V.~E.,  2020, in preparation

\bibitem[\protect\citeauthoryear{White \& Rees}{White \&
  Rees}{1978}]{White1978}
White S.~D.,  Rees M.~J.,  1978, Monthly Notices of the Royal Astronomical
  Society, 183, 341

\bibitem[\protect\citeauthoryear{Wong, Suyu, Chen, Rusu, Millon, Sluse, Bonvin
  \& Fassnacht}{Wong et~al.}{2020}]{Wong2020}
Wong K.~C.,  Suyu S.~H.,  Chen G. C.-F.,  Rusu C.~E.,  Millon M.,  Sluse D.,
  Bonvin V.,    Fassnacht C.~D.,  2020, Monthly Notices of the Royal
  Astronomical Society

\bibitem[\protect\citeauthoryear{Zhang, Cisse, Dauphin \& Lopez-Paz}{Zhang
  et~al.}{2018}]{zhang2018mixup}
Zhang H.,  Cisse M.,  Dauphin Y.~N.,    Lopez-Paz D.,  2018, in International
  Conference on Learning Representations {mixup: Beyond Empirical Risk
  Minimization}

\end{thebibliography}


\bsp	
\label{lastpage}
\end{document}